\documentclass[10pt,journal]{IEEEtran}
\usepackage{amssymb}
\usepackage{amsmath}
\usepackage{amsfonts}
\usepackage{epsfig}
\usepackage{graphicx}

\begin{document}

\title{Signal processing on graphs: Transforms and tomograms}
\date{ }
\author{R. Vilela Mendes \thanks{%
CMAF and IPFN, Univ. Lisbon, Av. Prof. Gama Pinto 2, 1649-003 Lisboa,
rvmendes@fc.ul.pt, rvilela.mendes@gmail.com}, Hugo C. Mendes \thanks{%
Instituto Portugu\^{e}s do Mar e da Atmosfera, Avenida Bras\'{\i}lia,
1300-598 Lisboa, hugart@gmail.com}, Tanya Ara\'{u}jo \thanks{%
ISEG, Technical University of Lisbon, Rua do Quelhas, 6 1200-781 Lisboa
Portugal, tanya@iseg.utl.pt}}
\maketitle

\begin{abstract}
Using projections on the (generalized) eigenvectors associated to matrices
that characterize the topological structure, several authors have
constructed generalizations of the Fourier transform on graphs. By exploring
mappings of the spectrum of these matrices we show how to construct more
general transforms, in particular wavelet-like transforms on graphs. For
time-series, tomograms, a generalization of the Radon transforms to
arbitrary pairs of non-commuting operators, are positive bilinear transforms
with a rigorous probabilistic interpretation which provide a full
characterization of the signals and are robust in the presence of noise.
Here the notion of tomogram transform is also extended to signals on
arbitrary graphs.
\end{abstract}


\section{Introduction}

\subsection{Signal transforms for time series: Linear, quasi-distributions
and tomograms}

The traditional field of signal processing deals mostly with the analysis of
time series. Signal processing of time series relies heavily on integral
transforms \cite{handbook} \cite{BWolf79}. Three types of transforms have
been used: linear, bilinear and tomograms. Among the linear transforms
Fourier and wavelets {are the most popular. The Fourier transform extracts
the frequency components of the signal and the wavelets its multiscale
nature. However, this is achieved at the expense of the time information, in
the sense that the time location of the frequency components and of the
scale features is lost in the process. This motivated the development of
bilinear transforms like the time-frequency Wigner-Ville \cite{Wigner32} 
\cite{Ville48} or the frequency-scale Bertrand \cite{BerBerJMP} \cite{Baran}
quasidistributions. The aim of the Wigner-Ville transform was to provide
joint information on the time-frequency plane, which is important because in
many applications (biomedical, seismic, radar, etc.) the signals are of
finite (sometimes very short) duration. However, the oscillating cross-terms
in the Wigner--Ville and other quasidistributions \cite{Cohen1} \cite{Cohen2}
\cite{Gabor} make the interpretation of the transformed signals a difficult
matter. Even when the average of the cross-terms is small, their amplitude
may be greater than the signal in time--frequency regions that carry no
physical information.}

{The difficulties with the physical interpretation of quasidistributions
arise from the fact that time and frequency (or frequency and scale) are
noncommutative operator pairs. Hence, a joint probability density can never
be defined.} Even in the case of positive quasiprobabilities like the
Husimi--Kano function \cite{Husimi} \cite{Kano}, an interpretation as a
joint probability distribution is also not possible because the two
arguments in the function are not simultaneously measurable random
variables. More recently, a new type of strictly positive bilinear transform
has been proposed \cite{MendesPLA} \cite{MankoJPA} \cite{RVMJRLR}, called a
tomogram, which is a generalization of the Radon transform \cite{Radon} to
arbitrary noncommutative pairs of operators. The Radon--Wigner transform 
\cite{Radon1} \cite{Radon2} is a particular case of the noncommutative
tomography technique. Being strictly positive probability densities, the
tomograms provide a full characterization of the signal and are robust in
the presence of noise.

A unified framework to characterize linear transforms, quasidistributions
and tomograms was developed in Ref.\cite{MankoJPA}. In short, considering a
signal $f(t)$ as a vectors $\left\vert f\right\rangle $ in a subspace $%
\mathcal{N}$ of a Hilbert space $\mathcal{H}$, a family of unitary operators 
$U\left( \alpha \right) =e^{iB\left( \alpha \right) }$ and a reference
vector $h$ in the dual $\mathcal{N}^{\ast }$ of $\mathcal{N}$ , a linear
transform like Fourier or wavelet is 
\begin{equation}
W_{f}^{(h)}(\alpha )=\langle U\left( \alpha \right) h\mid f\rangle
\label{1.1}
\end{equation}%
and a quasidistribution is 
\begin{equation}
Q_{f}(\alpha )=\langle U\left( \alpha \right) f\mid f\rangle  \label{1.2}
\end{equation}

To define the tomogram let, in the unitary operator $U\left( \alpha \right)
=e^{iB\left( \alpha \right) }$, $B\left( \alpha \right) $ have the spectral
decomposition $B\left( \alpha \right) =\int XP\left( X\right) dX$. Then%
\begin{equation*}
P\left( X\right) \circeq \left\vert X\right\rangle \left\langle X\right\vert
\end{equation*}%
denotes the projector on the (generalized) eigenvector $\left\langle
X\right\vert \in \mathcal{N}^{\ast }$ of $B\left( \alpha \right) $. The
tomogram is 
\begin{equation}
M_{f}^{(B)}(X)=\left\langle f\right\vert P\left( X\right) \left\vert
f\right\rangle =\left\langle f\right\vert \left. X\right\rangle \left\langle
X\right. \left\vert f\right\rangle =\left\vert \left\langle X\right.
\left\vert f\right\rangle \right\vert ^{2}  \label{1.3}
\end{equation}

The tomogram $M_{f}^{(B)}(X)$ is the squared amplitude of the projection of
the signal $\left\vert f\right\rangle \in \mathcal{N}$ on the eigenvector $%
\left\langle X\right\vert \in \mathcal{N}^{\ast }$ of the operator $B\left(
\alpha \right) $. Therefore it is positive. For normalized $\mid f\rangle $, 
\begin{equation*}
\langle f\mid f\rangle =1
\end{equation*}%
the tomogram is normalized 
\begin{equation}
\int M_{f}^{(B)}\left( X\right) \,dX=1  \label{1.5}
\end{equation}%
and may be interpreted as a probability distribution on the set of
generalized eigenvalues of $B\left( \alpha \right) $, that is, as the
probability distribution for the random variable $X$ corresponding to the
observable defined by the operator $B\left( \alpha \right) $.

For example, if the unitary $U\left( \alpha \right) $ is generated by $%
B_{F}\left( \overrightarrow{\alpha }\right) =\alpha _{1}t+i\alpha _{2}\frac{d%
}{dt}$ and $h$ is a (generalized) eigenvector of the time-translation
operator, the linear transform $W_{f}^{(h)}(\alpha )$ is the Fourier
transform. For the same $B_{F}\left( \overrightarrow{\alpha }\right) $%
\textbf{,} the quasi-distribution\textbf{\ }$Q_{f}(\alpha )$ is the
ambiguity function and the Wigner--Ville transform {\cite{Wigner32} \cite%
{Ville48}} is the quasi-distribution $Q_{f}(\alpha )$ for the following $B-$%
operator 
\begin{equation}
B^{(WV)}(\alpha _{1},\alpha _{2})=-i2\alpha _{1}\frac{d}{dt}-2\alpha _{2}t+%
\frac{\pi \left( t^{2}-\frac{d^{2}}{dt^{2}}-1\right) }{2}\,  \label{1.7}
\end{equation}

The wavelet transform is $W_{f}^{(h)}(\alpha )$ for $B_{W}\left( 
\overrightarrow{\alpha }\right) =\alpha _{1}D+i\alpha _{2}\frac{d}{dt}$, $D$
being the dilation operator $D=-\frac{1}{2}\left( it\frac{d}{dt}+i\frac{d}{dt%
}t\right) $. The wavelets $h_{s,\,\tau }(t)$ are kernel functions generated
from a basic wavelet $h(\tau )$ by means of a translation and a rescaling $%
(-\infty <\tau <\infty ,$ $s>0)$: 
\begin{equation}
h_{s,\,\tau }(t)=\frac{1}{\sqrt{s}}\,h\left( \frac{t-\tau }{s}\right)
\label{1.8}
\end{equation}%
using the operator 
\begin{equation}
U^{(A)}(\tau ,s)=\exp (i\tau \hat{\omega})\exp (i\log \,sD),  \label{1.9}
\end{equation}%
\begin{equation}
h_{s,\tau }(t)=U^{(A)\dagger }(\tau ,s)h(t).  \label{1.10}
\end{equation}

The Bertrand transform \cite{BerBerJMP} \cite{Baran} is the
quasi-distribution $Q_{f}(\alpha )$ for $B_{W}$. Linear, bilinear and
tomogram transforms are related to one another \cite{MankoJPA}.

As shown before, tomograms are obtained from projections on the eigenstates
of the $B$ operators. These operators may be linear combinations of
different (commuting or noncommuting) operators, 
\begin{equation*}
B=\mu O_{1}+\nu O_{2}
\end{equation*}%
meaning that the tomogram explores the signal along lines in the plane $%
\left( O_{1},O_{2}\right) $. For example for 
\begin{equation*}
B\left( \mu ,\nu \right) =\mu t+\nu \omega =\mu t+i\nu \frac{d}{dt}
\end{equation*}%
the tomogram is the expectation value of a projection operator with support
on a line in the time--frequency plane 
\begin{equation}
X=\mu t+\nu \omega  \label{3.2}
\end{equation}%
Therefore, $M_{f}^{(S)}\left( X,\mu ,\nu \right) $ is the marginal
distribution of the variable $X$ along this line in the time--frequency
plane. The line is rotated and rescaled when one changes the parameters $\mu 
$ and $\nu $. In this way, the whole time--frequency plane is sampled and
the tomographic transform contains all the information on the signal. The
probabilistic nature of the tomogram implies that, in contrast with
quasi-distributions, the information thus obtained is robust and
unambiguous. Tomograms associated to linear combinations of time with the
generators of the conformal group ($i\nu \frac{d}{dt};i\left( t\frac{d}{dt}+%
\frac{1}{2}\right) ;i\left( t^{2}\frac{d}{dt}+t\right) $) and several other
known operators have been explored \cite{RVMJRLR}.

By providing a robust extraction of compound signal features, tomograms have
been useful in denoising, component separation and structure identification 
\cite{reflecto1} \cite{reflecto2} \cite{reflecto3} \cite{Clairet} \cite%
{Aguirre} \cite{Aguirre2}.

\subsection{Signals on graphs}

Social and economic networks, information networks, power grids, biological
networks, etc. generate large sets of raw data from which, in general, only
a detailed analysis may extract useful information. A first step is the
construction of the appropriate signal transforms.

From the graph point of view a time series is a signal on a one-dimensional
directed graph with vertices labelled by the times $\left(
t_{0},t_{1},t_{2},\cdots \right) $ and the edges connecting $t_{k+1}$ to $%
t_{k}$. That is, the adjacency matrix $\mathbf{A}$ of a time series is, in
general%
\begin{equation}
\mathbf{A}=\left( 
\begin{array}{ccccc}
0 & 0 & 0 & 0 & \cdots \\ 
1 & 0 & 0 & 0 & \cdots \\ 
0 & 1 & 0 & 0 & \cdots \\ 
0 & 0 & 1 & 0 & \cdots \\ 
\vdots & \vdots & \vdots & \vdots & \cdots%
\end{array}%
\right)  \label{I.1}
\end{equation}%
or, for a time-periodic signal%
\begin{equation}
\mathbf{A}=\left( 
\begin{array}{cccccc}
0 & 0 & 0 & \cdots & 0 & 1 \\ 
1 & 0 & 0 & \cdots & 0 & 0 \\ 
0 & 1 & 0 & \cdots & 0 & 0 \\ 
0 & 0 & 1 & \cdots & 0 & 0 \\ 
\vdots & \vdots & \vdots & \vdots & 0 & 0 \\ 
0 & 0 & 0 & 0 & 1 & 0%
\end{array}%
\right)  \label{I.2}
\end{equation}%
As discussed before, signal transforms for a time series are projections on
a set of eigenvectors of some linear operator. These operators are not
arbitrary, but chosen to extract particular features of the signal that is
being analyzed. The Fourier transform looks for periodic features, wavelets
for multiscale features, etc. Likewise, useful information from signals on
arbitrary graphs may be obtained from projections on sets of vectors
associated to suitably chosen linear operators. For the time-periodic
signal, it is easy to see that the discrete Fourier transform is the
projection on the eigenvectors of the adjacency matrix (\ref{I.2}).
Therefore one may define the graph Fourier transform for an arbitrary graph
as the projection on the eigenvectors (or on the generalized eigenvectors of
the Jordan decomposition) of the adjacency matrix. This was the point of
view taken by some authors \cite{Moura1} to develop a theory of discrete
signal processing on graphs. However this choice is not unique because, for
the time series network other matrices have the same spectrum, for example
the Laplacian matrix%
\begin{equation*}
\mathbf{L}=\mathbf{D-A}
\end{equation*}%
$\mathbf{D}$ being the degree matrix, which for the time series is the
identity. Hence the graph Fourier transform might as well be defined as a
projection on the generalized eigenvectors of the Laplacian matrix \cite%
{Miller} \cite{Shuman}. This operator point of view allows not only to
generalize the notion of transforms but also the notions of filtering and
other general linear operations on graph signals.

\section{Signal transforms and tomograms on graphs}

Here a generalization of the notions of linear transform and tomogram for
signals on graphs will be developed. Generalization of the notion of
bilinear transform will not be dealt with because, already for time series,
it leads to difficult interpretation problems.

\subsection{Graph transforms}

Let $G=(\mathcal{V},\mathbf{A})$ be a graph, with $\mathcal{V}%
=\{v_{0},\ldots ,v_{N-1}\}$ the set of vertices and $\mathbf{A}$ the
weighted adjacency matrix. Each matrix element $\mathbf{A}_{n,m}$ is the
weight of a directed edge from $v_{m}$ to $v_{n}$ which can take arbitrary
real or complex values. $\mathcal{N}_{n}=\{m\mid \mathbf{A}_{n,m}\neq 0\}$
is the neighborhood of $v_{n}$ and a graph signal is a map $\mathbf{f}%
=\left\{ f_{n}\right\} $ from the set $\mathcal{V}$ of vertices into the set
of complex numbers $\mathbb{C}$, each element $f_{n}$ being indexed by the
vertex $v_{n}$.

Other useful matrices are:

- \textit{The degree matrix }$\mathbf{D}$\textit{:} a diagonal matrix
listing the degree of the vertices

- \textit{The Laplacian matrix:} $\mathbf{L}=\mathbf{D}-\mathbf{A}$

- \textit{The symmetrically normalized Laplacian matrix:} $\mathbf{L}%
^{\prime }\mathbf{=D}^{-\frac{1}{2}}\mathbf{LD}^{-\frac{1}{2}}$

- \textit{The random walk matrix:} $\mathbf{W}=\mathbf{AD}^{-1}$

- \textit{The lazy random walk matrix:} $\mathbf{W}^{\prime }=\left( \mathbf{%
I}+\mathbf{AD}^{-1}\right) /2$

- \textit{The incidence matrix }$\nabla $\textit{:} is the $m\times N$
matrix ($m$=no. edges, $N$=no. of vertices) given by%
\begin{equation*}
\nabla _{e,v}=\left\{ 
\begin{array}{cccc}
1 & \text{if} & e=\left( v,w\right) & \text{and }v<w \\ 
-1 & \text{if} & e=\left( v,w\right) & \text{and }v>w \\ 
0 &  & \text{otherwise} & 
\end{array}%
\right.
\end{equation*}

- \textit{The edge adjacency matrix:} is a $m\times m$ matrix determined by
the adjacencies of edges%
\begin{equation*}
^{e}\mathbf{A}_{i,j}\mathbf{=}\left\{ 
\begin{array}{ll}
1 & \text{if edges i and j are adjacent} \\ 
0 & \text{otherwise}%
\end{array}%
\right.
\end{equation*}

These matrices have been used in the past mostly to characterize the
topological structure of networks, for vertex clustering, detection of
communities, etc. \cite{VonLuxburg} \cite{Florent} \cite{Newman1} \cite%
{Hashimoto} \cite{Newman2}. Here they will be considered as operators which
generate a set of (generalized) eigenvectors to project the signals on
graphs.\bigskip

\textbf{Fourier-like transforms}

Denote any one of these matrices by $\mathbf{M}$. The matrices $\mathbf{M}$
act on the space of graph signals by%
\begin{equation}
f\rightarrow \widetilde{f}_{n}=\sum_{m}\mathbf{M}_{n,m}f_{m}=\sum_{m\in 
\mathcal{N}_{n}}\mathbf{M}_{n,m}f_{m}.  \label{5.1}
\end{equation}%
When the matrix $\mathbf{M}$ is the adjacency matrix this operation
generalizes the notion of time shift, when time sequences are looked at as
forward-connected graphs.

For many real-world datasets the matrices $\mathbf{M}$ are not
diagonalizable. In those cases, to obtain a suitable set of expansion
vectors one may either use the symmetric combinations $\mathbf{MM}^{T}$ and $%
\mathbf{M}^{T}\mathbf{M}$ to generate an expansion basis or, alternatively,
use the block-diagonal Jordan decomposition of $\mathbf{M}$.%
\begin{equation}
\mathbf{M}=VJV^{-1}  \label{5.2}
\end{equation}%
\begin{equation}
J=%
\begin{pmatrix}
J_{R_{0,0}}(\lambda _{0}) &  &  \\ 
& \ddots &  \\ 
&  & J_{R_{M-1,D_{M-1}}}(\lambda _{M-1})%
\end{pmatrix}
\label{5.3}
\end{equation}%
with Jordan blocks associated to the eigenvalues of $\mathbf{M}$%
\begin{equation}
J_{r_{m,d}}(\lambda _{m})=%
\begin{pmatrix}
\lambda _{m} & 1 &  &  \\ 
& \lambda _{m} & \ddots &  \\ 
&  & \ddots & 1 \\ 
&  &  & \lambda _{m}%
\end{pmatrix}
\label{5.4}
\end{equation}%
The columns of the matrix $V$, that brings $\mathbf{M}$ to its Jordan normal
form, are the eigenvectors%
\begin{equation}
(\mathbf{M}-\lambda _{m}\mathbf{1})\mathbf{v}_{m,d,0}=0  \label{5.5}
\end{equation}%
and the generalized eigenvectors of the Jordan chain%
\begin{equation}
(\mathbf{M}-\lambda _{m}\mathbf{1})\mathbf{v}_{m,d,r}=\mathbf{v}_{m,d,r-1}
\label{5.6}
\end{equation}%
of $\mathbf{M}$. These vectors may then be used to project the signals on
the graph and, considering the graph signal $\mathbf{f}$ as a column vector,
the $\mathbf{M-}$transform is then%
\begin{equation}
\mathbf{\widehat{f}}=V^{-1}\mathbf{f}  \label{5.7}
\end{equation}%
with inverse transform%
\begin{equation}
\mathbf{f}=V\mathbf{\widehat{f}}  \label{5.8}
\end{equation}

The problem with this decomposition lies in the fact that in general the set
of generalized eigenvectors do not form an orthogonal basis. Therefore it is
sometimes more convenient to use $\mathbf{MM}^{T}$ and $\mathbf{M}^{T}%
\mathbf{M}$ to generate the expansion basis, leading to what we will call
the $\mathbf{MM}^{T}-$ or $\mathbf{M}^{T}\mathbf{M-}$transform.\bigskip

\textbf{Wavelet-like transforms}

The definition of wavelet-like transforms for graphs requires a more
elaborate construction. For time series the affine wavelets use, in Eq. (\ref%
{1.1}), an operator $U\left( \alpha \right) $ consisting of the product of a
translation and a scale transformation which acts on a fixed reference
signal (the mother wavelet $h_{0}\left( t\right) $), namely%
\begin{equation}
h_{s,a}\left( t\right) =U\left( s,a\right) h_{0}\left( t\right) =e^{\log
s\left( t\frac{d}{dt}+\frac{1}{2}\right) }e^{a\frac{d}{dt}}h_{0}\left(
t\right) =\sqrt{s}h_{0}\left( st+a\right)  \label{w.1}
\end{equation}%
Translation in the graph is easily generalized but it is not obvious how to
generalize scale transformations. Hence we rewrite the wavelet transform in
frequency space obtaining%
\begin{eqnarray}
f\left( a,s\right) &=&\int dth_{s,a}^{\ast }\left( t\right) f\left( t\right)
=\int dt\left( e^{\log s\left( t\frac{d}{dt}+\frac{1}{2}\right) }e^{a\frac{d%
}{dt}}h_{0}^{\ast }\left( t\right) \right) f\left( t\right)  \notag \\
&=&\int d\omega \frac{e^{-i\frac{\omega }{s}a}}{\sqrt{s}}\widehat{h_{0}}%
^{\ast }\left( \frac{\omega }{s}\right) \widehat{f}\left( \omega \right)
\label{w.2}
\end{eqnarray}%
$\widehat{h_{0}}$ and $\widehat{f}$ denoting the Fourier transforms of the
mother wavelet and of the signal. One sees that the wavelet transform is
represented as a sum over the Fourier spectrum $\Omega $ with the argument
of the mother wavelet shifted from $\omega $ to $\frac{\omega }{s}$. The
mapping $\omega \in \Omega \rightarrow \frac{\omega }{s}\in \Omega $ is a
one-to-one onto mapping of $\Omega $ in $\Omega $. Therefore the natural
generalization of the wavelet transform for graphs may be defined as a
similar sum, with the spectrum label shift being one of the possible
one-to-one onto mappings of the spectrum of the adjacency matrix (or of the
Laplacian matrix).

Consider the Fourier-like transform on graphs and its inverse%
\begin{eqnarray}
f\left( i\right) &=&\sum_{\eta }\widehat{f}\left( \eta \right) \chi _{\eta
}\left( i\right)  \notag \\
\widehat{f}\left( \eta \right) &=&\sum_{i}\chi _{\eta }\left( i\right)
f\left( i\right)  \label{w.3}
\end{eqnarray}%
$\chi _{\eta }\left( i\right) $ being an eigenvector of $A$ or $L$ (or a
generalized eigenvector or an eigenvector of $A^{T}A$ or $L^{T}L$) and a
localized "mother wavelet"%
\begin{equation}
h^{(k)}\left( i\right) =\delta _{k,i}  \label{w.4}
\end{equation}%
Then the wavelet-like transform on graphs would be%
\begin{equation}
f\left( a,\widetilde{s}\right) =\sum_{\eta }\chi _{\widetilde{s}\left( \eta
\right) }\left( k+a\right) \widehat{f}\left( \eta \right)  \label{w.5}
\end{equation}%
$\widetilde{s}\left( \eta \right) $ is not $\eta \rightarrow \frac{\eta }{s}$
because in general $\frac{\eta }{s}$ is not in $\Omega $. $\widetilde{s}$
would be a mapping in the set $\mathcal{S}$ of the possible one-to-one onto
mappings of $\Omega $, $\widetilde{s}\in \mathcal{S}$.

The inverse wavelet transform is%
\begin{equation}
\widehat{f}\left( \eta \right) =\frac{1}{\#\mathcal{S}}\sum_{a,\overset{%
\symbol{126}}{s}}\chi _{\widetilde{s}\left( \eta \right) }\left( a\right)
f\left( a,\overset{\symbol{126}}{s}\right)  \label{w.6}
\end{equation}

Hammond, Vandergheynst and Gribonval \cite{Hammond} have also attempted to
generalize the notion of wavelet transform to graph signals. However,
instead of the sum with the shifted arguments in the spectrum, they simply
use a $\eta -$dependent weight on the sum with both the signal component $%
\overset{\curlywedge }{f}\left( \eta \right) $ and the eigenvector $\chi
_{\eta }$ associated to the same spectral value $\eta $. Therefore their
construction is more in the spirit of a superposition of Fourier-like
transforms than of a wavelet transform.

A more general transform would be%
\begin{equation}
f\left( a,C\right) =\sum_{\eta ,\eta ^{\prime }}C\left( \eta ,\eta ^{\prime
}\right) \chi _{\eta ^{\prime }}\left( a\right) \overset{\curlywedge }{f}%
\left( \eta \right)  \label{w.7}
\end{equation}%
For comparison with the time series case, this last construction would be
similar to the case of the "conformal wavelets" generated by $e^{\alpha
\left( t^{2}\frac{d}{dt}+t\right) }e^{a\frac{d}{dt}}h_{0}\left( t\right) $.

\subsection{Graph tomograms}

So far signals on graphs have been described either as vectors on vertex
space or as projections of these vectors on the generalized eigenvectors of
a particular matrix $\mathbf{M}$. Each particular matrix emphasizes a
specific topological property of the graph. Tomograms attempt to obtain
information about more than one property by projecting on the generalized
eigenvectors of a matrix that interpolates between two distinct matrices $%
\mathbf{M}_{1}$ and $\mathbf{M}_{2}$. This parallels what for time series is
achieved, for example, by the time-frequency tomogram.

When the vertex space has a meaningful physical interpretation it is useful
to interpolate between one of the matrices $\mathbf{M}$ listed before and
the matrix for which the vertex signal corresponds to a projection on its
eigenvectors. For a graph with $N$ vertices, the vectors on vertex space may
be considered as projections on the eigenvectors of a \textit{vertex operator%
}%
\begin{equation}
\mathbf{T}=\left( 
\begin{array}{ccccc}
1 & 0 & 0 & \vdots & 0 \\ 
0 & e^{i\frac{2\pi }{N}} & 0 & \vdots & 0 \\ 
0 & 0 & e^{i\frac{2\pi }{N}\times 2} & \vdots & 0 \\ 
\cdots & \cdots & \cdots & \ddots & \vdots \\ 
0 & 0 & 0 & \cdots & e^{i\frac{2\pi }{N}\times \left( N-1\right) }%
\end{array}%
\right)  \label{5.9}
\end{equation}%
Therefore the construction of a tomogram for graph signals would amount to
finding an operator that interpolates between $T$ and $\mathbf{A}$. A
solution could be the family of operators%
\begin{equation}
B_{\alpha }=\left( 1-\alpha \right) \mathbf{T}+\alpha \mathbf{M}
\label{5.10}
\end{equation}%
with $\alpha $ varying between $0$ and $1$, the tomogram being obtained by
the projections of the signal on the eigenvectors of $B_{\alpha }$.

If $\mathbf{M}$ is the adjacency matrix $\mathbf{A}$, this construction,
interpolating between $\mathbf{A}$ and the vertex operator $\mathbf{T}$, is
for graphs, the analog of the time-frequency tomogram.

If the ordering of the vertices is arbitrary, the vertex operator has no
special meaning and it is more useful to construct tomograms using two of
the listed $\mathbf{M}$ matrices which, by construction, already contain
meaningful information on the graph.

As discussed before, the reason why time and frequency cannot be
simultaneously specified is because they correspond to a pair of
non-commuting operators. This is the reason why bilinear transforms, like
Wigner-Ville, are unreliable and it is also the main motivation for using
tomogram transforms. In graphs also, the vertex description and the
adjacency matrix projection are also incompatible specifications, because in
general the $\mathbf{T}$ and $\mathbf{A}$ (or $\mathbf{L}$) matrices do not
commute. It is in this sense, that, as recently stated \cite{Agaskar}, there
is an uncertainty principle for graphs, that is, a fundamental trade-off
between a signal localization on the graph and on its spectral domain.

\subsection{Tomograms and dynamics}

The graph tomogram, as defined above, is appropriate for the study of a
static network signal\footnote{%
Likewise, the usual time-frequency tomogram may be looked upon as a static
description of the whole time history of the system.}. If during the time
evolution the graph structure stays the same, the time series associated to
each vertex may simply be projected on the (generalized) eigenvectors as in
the scalar case. However if the graph itself changes in time a more general
framework must be used.

Consider a graph signal that evolves in (discrete) time. The corresponding
graph would be, for each time $t$, a regular graph and each one of these
graphs is forward-connected to the graph of the subsequent time. A vertex $%
\nu _{n}\left( t\right) $ at time $t$ connects to the vertex $\nu _{n}\left(
t+1\right) $ at time $t+1$. This construction accommodates the possible
disappearance of vertices. In that case such vertex $\nu _{n}\left( t\right) 
$ would not have any forward edges.

The construction of the $\mathbf{M-}$transforms and the graph tomograms will
then proceed as before for the global adjacency matrix. To have a feeling
for the kind of eigenvectors obtained for such adjacency matrices, consider
a simple case of a finite-vertex circle graph with $N$ vertices
symmetrically connected to nearest-neighbors and forward connected in
periodic time with $\tau $ time steps. Then, at each time $t$, the
subadjacency matrix $\mathbf{A}\left( t\right) $ is%
\begin{equation}
\mathbf{A}\left( t\right) =\left( 
\begin{array}{cccccc}
0 & 1 & 0 & 0 & \vdots & 1 \\ 
1 & 0 & 1 & 0 & \vdots & 0 \\ 
0 & 1 & 0 & 1 & \vdots & 0 \\ 
0 & 0 & 1 & 0 & \vdots & 0 \\ 
\cdots & \cdots & \cdots & \cdots & \ddots & \vdots \\ 
0 & 0 & 0 & \cdots & 1 & 0%
\end{array}%
\right)  \label{5.17}
\end{equation}%
Let, for definiteness and notational simplicity, $N=\tau =3$. Then the
global $9\times 9$ adjacency matrix is%
\begin{equation}
\mathbf{A}=\left( 
\begin{array}{ccccccccc}
0 & 1 & 1 & 0 & 0 & 0 & 1 & 0 & 0 \\ 
1 & 0 & 1 & 0 & 0 & 0 & 0 & 1 & 0 \\ 
0 & 1 & 0 & 0 & 0 & 0 & 0 & 0 & 1 \\ 
1 & 0 & 0 & 0 & 1 & 1 & 0 & 0 & 0 \\ 
0 & 1 & 0 & 1 & 0 & 1 & 0 & 0 & 0 \\ 
0 & 0 & 1 & 0 & 1 & 0 & 0 & 0 & 0 \\ 
0 & 0 & 0 & 1 & 0 & 0 & 0 & 1 & 1 \\ 
0 & 0 & 0 & 0 & 1 & 0 & 1 & 0 & 1 \\ 
0 & 0 & 0 & 0 & 0 & 1 & 0 & 1 & 0%
\end{array}%
\right)  \label{5.18}
\end{equation}%
This matrix is a tensor product of matrices 
\begin{equation*}
\mathbf{A}=\left( 
\begin{array}{ccc}
0 & 1 & 1 \\ 
1 & 0 & 1 \\ 
0 & 1 & 0%
\end{array}%
\right) \otimes \left( 
\begin{array}{ccc}
0 & 0 & 1 \\ 
1 & 0 & 0 \\ 
0 & 1 & 0%
\end{array}%
\right)
\end{equation*}%
with eigenvalues, respectively%
\begin{equation}
\left( 
\begin{array}{c}
-1 \\ 
\frac{1}{2}\left( 1+\sqrt{5}\right) \\ 
\frac{1}{2}\left( 1-\sqrt{5}\right)%
\end{array}%
\right) \hspace{0.5cm}\text{and\hspace{0.5cm}}\left( 
\begin{array}{c}
1 \\ 
e^{i\frac{2\pi }{3}} \\ 
e^{i\frac{4\pi }{3}}%
\end{array}%
\right)  \label{5.19}
\end{equation}%
the eigenvectors of $\mathbf{A}$ being the tensor products of the
eigenvectors of these matrices. The "Fourier" transform of a dynamical graph
signal will be the projection on these $9-$dimensional eigenvectors.

For the construction of the tomogram, the vertex operator $\mathbf{T}$, as
in (\ref{5.9}), is%
\begin{equation*}
\mathbf{T}=\left( 
\begin{array}{ccc}
\mathbf{T}^{(3)} & \mathbf{0} & \mathbf{0} \\ 
\mathbf{0} & \mathbf{T}^{(3)} & \mathbf{0} \\ 
\mathbf{0} & \mathbf{0} & \mathbf{T}^{(3)}%
\end{array}%
\right) 
\end{equation*}%
where $\mathbf{T}^{(3)}$ is the $3\times 3$ matrix $%
\begin{array}{ccc}
1 & 0 & 0 \\ 
0 & e^{i\frac{2\pi }{3}} & 0 \\ 
0 & 0 & e^{i\frac{2\pi }{3}\times 2}%
\end{array}%
$.

\section{Illustrative examples}

In this section we present some examples of the use of graph transforms and
graph tomograms. The detailed economic and biological implications of the
examples are beyond the scope of this paper. The examples are included only
as an illustration of the concepts and also as a guide on how the graph
formulation may be a powerful tool to analyze multivariate time series.

\subsection{A market network}

An important problem in the design of portfolios or ETF's is the
classification of the dynamical behavior of the trading values of market
products. Identifying clusters of products with similar dynamical behavior
allows the design of simpler portfolios, by the selection of representative
elements in each cluster. Here we analyze the daily closing equity prices of 
$301$ companies in the SP500 throughout the $250$ trading days of 2012. For
the purpose of the calculations the companies are ordered by sectors. The
company ticker symbols and GICS sector codes are listed in the Appendix.

From the daily returns%
\begin{equation}
r\left( t\right) =\log S\left( t\right) -\log S\left( t-1\right)  \label{6.1}
\end{equation}%
$S\left( t\right) $ being the closing price at day $t$, one computes a
dynamical distance between the company stocks $i$ and $j$ by%
\begin{equation}
d_{ij}=\sqrt{\sum_{t=1}^{250}\left( r_{i}\left( t\right) -r_{j}\left(
t\right) \right) ^{2}}  \label{6.2}
\end{equation}%
the sum being over the $250$ trading days in 2012.

Now one computes the smallest non-zero $d_{ij}$ ($d_{\min }$)\ and an
adjacency matrix $A$ with matrix elements $A_{ij}$ may be defined either by%
\begin{equation}
A_{ij}^{\#}=\frac{d_{\min }}{d_{ij}}\left( 1-\delta _{ij}\right)  \label{6.3}
\end{equation}%
or%
\begin{equation}
A_{ij}^{\left( \beta \right) }=\left( 1-\delta _{ij}\right) \exp \left(
-\beta \left( d_{ij}-d_{\min }\right) \right)  \label{6.4}
\end{equation}%
The second form is sometimes the most convenient one because, by varying $%
\beta $, one obtains a multiscale analysis of the dynamical similarities of
the companies. In Figs. 1 and 2 we show the color-coded adjacency matrices $%
A $ and $A^{\left( \beta =2\right) }$ that are obtained. One sees that the $%
A^{\left( \beta =2\right) }-$adjacency matrix provides a more detailed
picture of the nature of correlations between the return behavior of these
equities. From inspection of this matrix one already sees that although the
strongest correlations are on the "utilities" sector (GICS code 55), many
other inter-sectors correlations do exist. The main purpose of the analysis
is precisely to identify sets of companies with similar return behavior.

\begin{figure}[htb]
\centering
\includegraphics[width=0.4\textwidth]{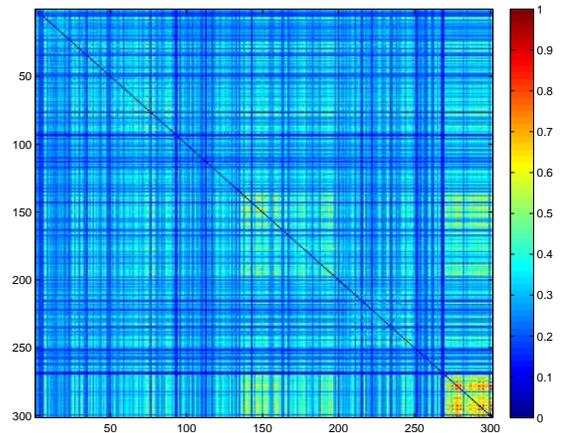}
\caption{Color-coded adjacency matrix $A^{\#}$ for the 301 companies}
\end{figure}

\begin{figure}[htb]
\centering
\includegraphics[width=0.4\textwidth]{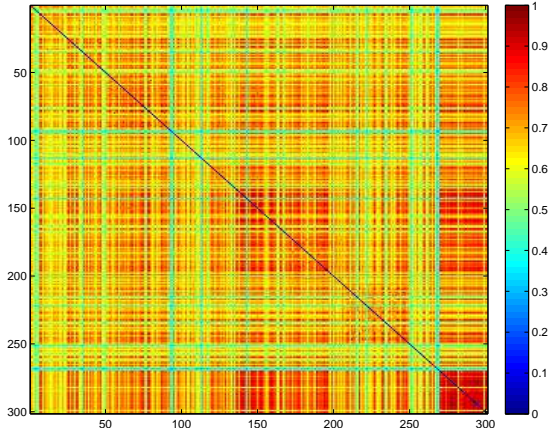}
\caption{Color-coded adjacency matrix $A^{\left( \protect\beta \right) }$, $%
\protect\beta =2$, for the 301 companies}
\end{figure}

For the remaining of our calculations we will use the $A^{\left( \beta
=2\right) }\circeq A$ as the adjacency matrix.

Now consider, as the signal on this graph, the yearly compound return%
\begin{equation}
R_{i}=\prod_{t=1}^{250}\left( 1+r_{i}\left( t\right) \right)  \label{6.5}
\end{equation}%
In Fig. 3 we compare the compound return $R_{i}$ of the companies with the
absolute value of the projections of $R_{i}-\left\langle R_{i}\right\rangle $
on the eigenvectors of the adjacency $A$ and the Laplacian $L=D-A$ matrices. 
$\left\langle R_{i}\right\rangle $ is the mean value of the compound
returns, which in this case was $1.1003$.

\begin{figure}[htb]
\centering
\includegraphics[width=0.4\textwidth]{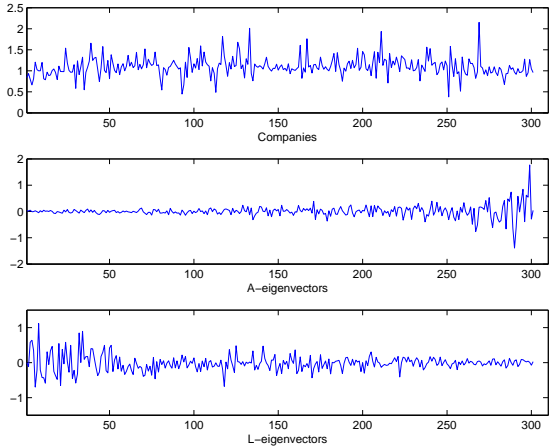}
\caption{The compound returns $R_{i}$ and the absolute values of the
projection of $R_{i}-\left\langle R_{i}\right\rangle $ on the eigenvectors
of the adjacency and Laplacian matrices}
\end{figure}

One sees that the projection on the $A-$eigenvectors (the $A-$transform) is
the one that provides a better information compression by selecting a
smaller number of dominant eigenvectors.

A standard spectral technique to find clusters in a graph is to look at the
lowest non-zero eigenvalues in the spectrum of the Laplacian matrix, the
corresponding eigenvectors leading (by K-means) to a division into clusters
that minimizes the RatioCut \cite{VonLuxburg}

\begin{equation*}
RatioCut\left( C_{1},...,C_{K}\right) =\frac{1}{2}\sum_{k=1}^{K}\frac{%
W\left( C_{k},\overline{C_{k}}\right) }{\left\vert C_{k}\right\vert }
\end{equation*}%
where $W\left( C_{k},\overline{C_{k}}\right) =\sum_{i\in C_{k},j\in 
\overline{C_{k}}}A_{ij}$, $\overline{C_{k}}$ is the complement of $C_{k}$
and $\left\vert C_{k}\right\vert $ is the number of elements in the cluster $%
C_{k}$.

From Fig.4, where we have plotted the eigenvalues of the Laplacian matrix,
one sees that in this case this criterium does not provide clear information
on the cluster properties of the market network.

\begin{figure}[htb]
\centering
\includegraphics[width=0.4\textwidth]{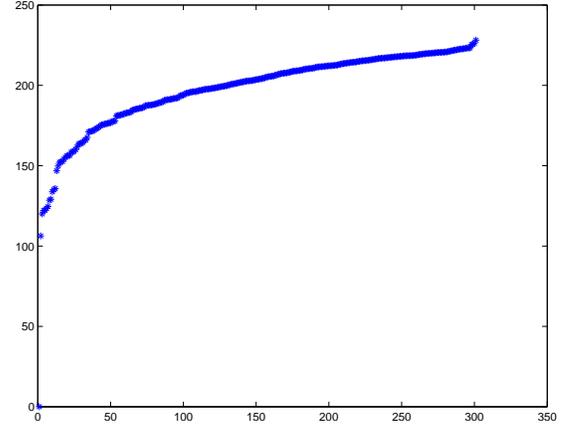}
\caption{Eigenvalues of the Laplacian matrix}
\end{figure}

How the companies organize themselves into groups with similar return
behavior is better understood by the examination of the $T-A$ tomogram
(Fig.5). The figure is a contour plot of the absolute value of the
projections of the compound return (Eq.\ref{6.5}) on the eigenvectors of $%
B_{\alpha }=\left( 1-\alpha \right) \mathbf{T}+\alpha \mathbf{A}^{\left(
2\right) }$.

\begin{figure}[htb]
\centering
\includegraphics[width=0.4\textwidth]{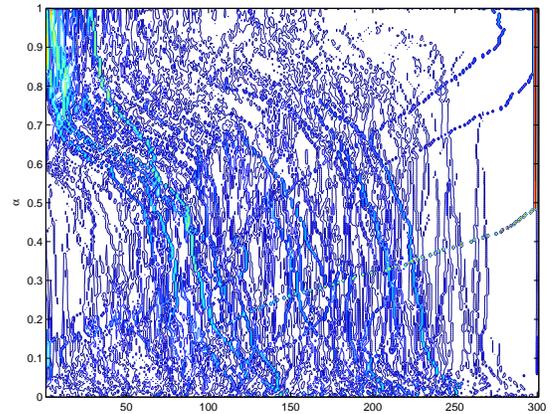}
\caption{The $T-A$ tomogram}
\end{figure}

One sees how, starting from the compound return signal at $\alpha =0$, the
contributions of the companies organize themselves into clusters on the way
to the final projection on the $\mathbf{A}-$eigenvectors (at $\alpha =1$).
The selection of clusters may be done by cutting the tomogram at diverse
levels and reconstructing the components of the signal. The tomogram has a
rigorous probabilistic interpretation and all the signal information is
contained at each $\alpha $ level. Therefore the signal components
(dynamical clusters) are reconstructed by linear combinations of the
eigenvectors around each peak with the coefficients taken from the tomogram.
As an example Fig. 6 shows the cut at $\alpha =0.85$ and Fig. 7 the
reconstruction of the signal components around three of its peaks.

\begin{figure}[htb]
\centering
\includegraphics[width=0.4\textwidth]{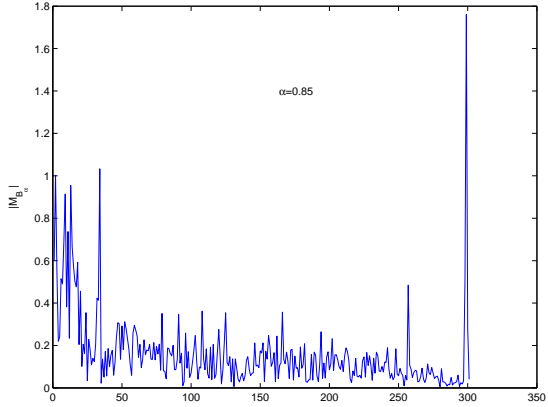}
\caption{The $T-A$ tomogram cut at $\protect\alpha =0.85$}
\end{figure}

\begin{figure}[htb]
\centering
\includegraphics[width=0.4\textwidth]{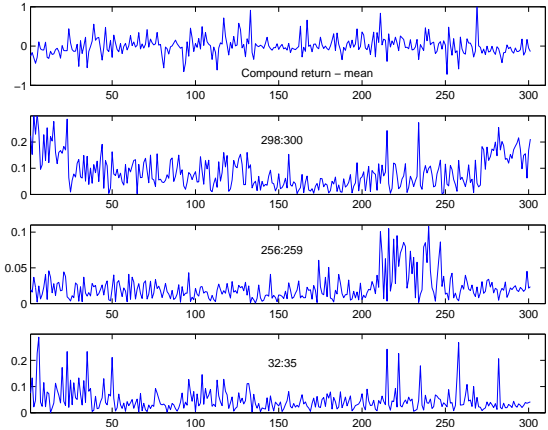}
\caption{The compound return and the absolute value of $R_{i}-\left\langle
R_{i}\right\rangle $ for three different clusters in the tomogram}
\end{figure}

One sees how these distinct dynamical clusters have important contributions
from different sectors. For example the last peak (components 298 to 300) is
dominated by companies both in the utilities and the energy sector.

\subsection{A trophic network}

In this example, to be studied in more detail elsewhere, we analyze a
biological network for which two types of information are available. It
concerns 12 fish species of the North Atlantic for which we have information
both on their trophic relations and on their biomass evolution in the period
1976-2013. These species were selected for the availability of a relatively
long biomass time series. The trophic relations, obtained from averaged
stomach sampling are displayed in Fig. 8 and in the color-coded adjacency
matrix $A_{\text{troph}}$ of Fig. 9.

\begin{figure}[htb]
\centering
\includegraphics[width=0.4\textwidth]{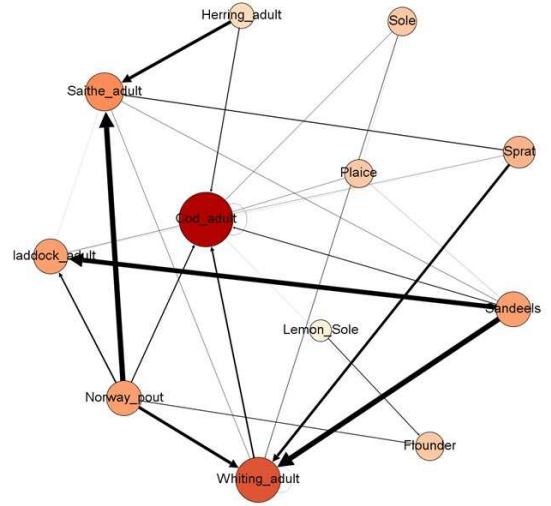}
\caption{North Sea Foodweb 12 species, directed and weighted}
\end{figure}

The ordered 12 species are: 1 = Cod adult; 2 = Whiting adult; 3 = Haddock
adult; 4 = Saithe adult; 5 = Norway pout; 6 = Herring adult; 7 = Sprat; 8 =
Sandeels; 9 = Plaice; 10 = Flounder; 11 = Sole; 12 = Lemon Sole.

\begin{figure}[htb]
\centering
\includegraphics[width=0.4\textwidth]{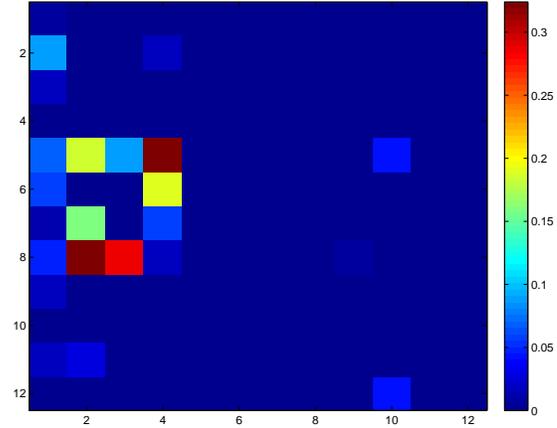}
\caption{Color-code trophic adjacency matrix $A_{\text{troph}}$ for the 12
fish species}
\end{figure}

Notice that in the $A_{\text{troph}}$ matrix the lines do not sum up to one,
because other species enter in the stomach data beyond the 12 considered
here.

On the other hand, considering, for each biomass time series $b\left(
t\right) $, the population growth rate as the most relevant variable \cite%
{Niwa}%
\begin{equation}
r\left( t\right) =\log \left( \frac{b\left( t\right) }{b\left( t-1\right) }%
\right)  \label{7.1}
\end{equation}%
we define the $\Delta -$ delay distance function%
\begin{equation}
d_{ij}^{\left( \Delta \right) }=\sqrt{\sum_{t=\Delta }^{38}\left(
r_{i}\left( t\right) -r_{j}\left( t-\Delta \right) \right) ^{2}}  \label{7.2}
\end{equation}%
The reason to consider time-delays for the growth rate distances is because
in a trophic network the biomass is related to the other species offspring
of previous years.

For each distance matrix, with elements $d_{ij}^{\left( \Delta \right) }$,
we find the smallest nonzero element ($d_{\min }^{\left( \Delta \right) }$)\
and define\ biomass delayed adjacency matrices as%
\begin{equation}
A_{ij}^{\left( \Delta \right) }=\frac{d_{\min }^{\left( \Delta \right) }}{%
d_{ij}^{\left( \Delta \right) }}  \label{7.3}
\end{equation}%
Fig. 10 displays the color-code one-year delayed biomass adjacency matrix.

\begin{figure}[htb]
\centering
\includegraphics[width=0.4\textwidth]{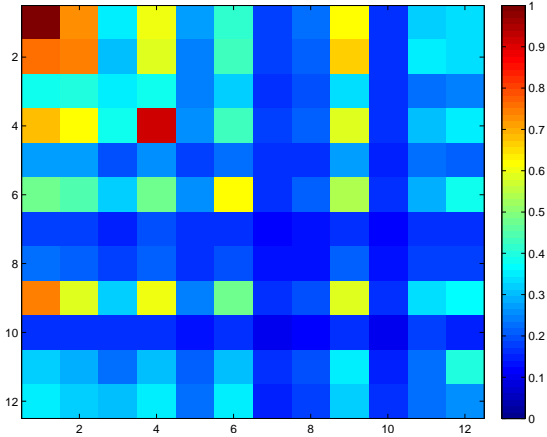}
\caption{Color-code one-year delayed biomass adjacency matrix}
\end{figure}

A simple inspection of Figs. 9 and 10 shows that the trophic and the biomass
data do not contain the same information, which is to be expected since the
biomass growth rate depends in many other factors besides predation. This is
better seen in Fig.11 where we have normalized to one each column in the
trophic matrix, and then compared the 28 nonzero elements with the
corresponding elements in the $A_{ij}^{\left( \Delta \right) }$ matrices
(also normalized to one).

\begin{figure}[htb]
\centering
\includegraphics[width=0.4\textwidth]{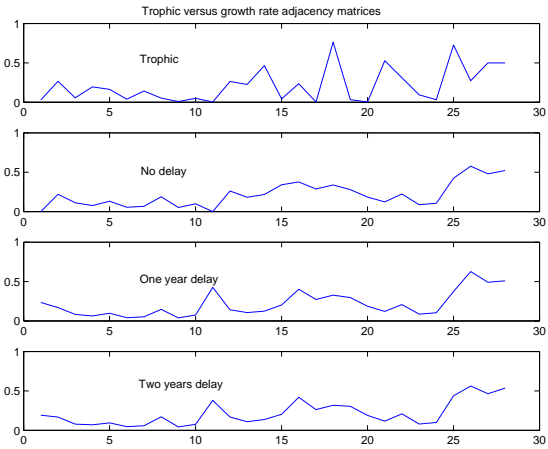}
\caption{Trophic versus growth rate adjacency matrices}
\end{figure}

Although some partial trends might be similar, the general conclusion is
that the biomass growth rate evolution seems to depend on many other
factors, different from the trophic relations of these 12 species.

In the remaining of this subsection we will use the one-year delayed biomass
growth rate and the tomographic analysis to exhibit the interspecies
correlations. Fig. 12 shows a contour plot of the tomogram corresponding to
the operator $B=\left( 1-\alpha \right) T+\alpha A^{\left( 1\right)
}A^{\left( 1\right) T}$. The signal that is projected on the eigenvectors of
this operator is $R_{i}-\left\langle R_{i}\right\rangle $, $R_{i}$ being the
compound growth rate over 36 years%
\begin{equation*}
R_{i}=\prod_{t=1}^{36}\left( 1+r_{i}\left( t\right) \right)
\end{equation*}%
The breaks that are observed in the contour plot result from the automatic
ordering of the eigenvectors by ascending eigenvalue values. They are of no
practical consequence, full information on the signal being kept at all $%
\alpha -$levels. One sees how, for $\alpha \neq 0$ the signal information is
compressed in a small number of eigenvectors.

\begin{figure}[htb]
\centering
\includegraphics[width=0.4\textwidth]{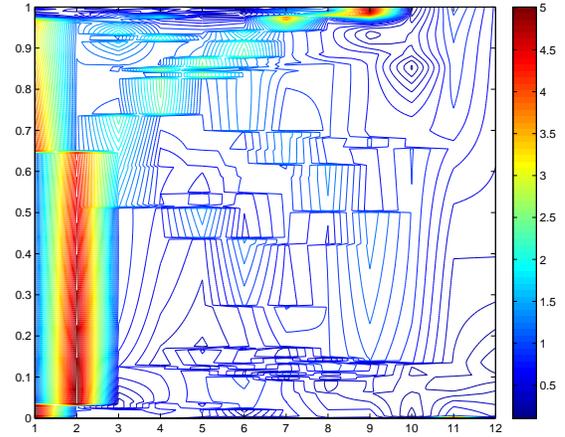}
\caption{The tomogram corresponding to the operator $B=\left( 1-\protect%
\alpha \right) T+\protect\alpha A^{\left( 1\right) }A^{\left( 1\right) T}$}
\end{figure}

As in the market network example, cutting the tomogram at intermediate $%
\alpha $ levels, clustering dependency of the species is obtained. This will
be reported in detail elsewhere. Here we only want to illustrate the use of
the graph tools that have developed and the promising power of the graph
formulation for the analysis of multivariate time series.\bigskip

{\LARGE Appendix}

\textbf{Ticker symbols and GICS sector codes of the SP500 companies used in
the example 3.1\bigskip }

APA(10); APC(10); BHI(10); CHK(10); CNX(10); COG(10); COP(10); CVX(10);
DNR(10); DO(10);\ DVN(10); FTI(10); HAL(10); HES(10); HP(10); MRO(10);
NBL(10); NOV(10); OXY(10); PXD(10); RDC(10); SLB(10); SWN(10); VLO(10);
WMB(10); XOM(10); AA(15); APD(15); ARG(15); ATI(15); BLL(15); BMY(15);
CAM(15); CF(15); CLF(15); DD(15); DOW(15); ECL(15); EMN(15); IFF(15);
IP(15); MON(15); MUR(15); NEM(15); NUE(15); PPG(15); PX(15); SEE(15);
VMC(15); X(15); APH(20); AVY(20); BA(20); CAT(20); CMI(20); CSX(20); DE(20);
DHR(20); DNB(20); DOV(20); EFX(20); EMR(20); ETN(20); FDX(20); FLR(20);
FLS(20); GD(20); GE(20); GWW(20); HON(20); IR(20); IRM(20); ITW(20);
LLL(20); LMT(20); LUV(20); MAS(20); MMM(20); NOC(20); NSC(20); PBI(20);
PH(20); PLL(20); R(20); RHI(20); ROK(20); RTN(20); TYC(20); UNP(20);
UTX(20); AN(25); AZO(25); BBY(25); BIG(25); CCE(25); COH(25); DFS(25);
DIS(25); DRI(25); F(25); FDO(25); GCI(25); GPC(25); GPS(25); HAR(25);
HD(25); HOG(25); HOT(25); HRB(25); IGT(25); IPG(25); JCI(25); JCP(25);
JWN(25); KMX(25); LEG(25); LEN(25); LOW(25); LTD(25); MCD(25); MHP(25);
NKE(25); NWL(25); OMC(25); PCP(25); SHW(25); SNA(25); SWK(25); TGT(25);
TIF(25); TJX(25); VFC(25); WHR(25); ADM(30); AVP(30); BFb(30); CAG(30);
CCL(30); CL(30); CLX(30); CPB(30); CVS(30); DF(30); DPS(30); EL(30);
GIS(30); HNZ(30); HRL(30); HSY(30); JEC(30); K(30); KMB(30); KO(30); KR(30);
LO(30); M(30); MJN(30); MO(30); PEP(30); PG(30); PM(30); SJM(30); STZ(30);
SWY(30); SYY(30); TAP(30); TSO(30); WAG(30); WMT(30); WPO(30); ABC(35);
ABT(35); AET(35); AGN(35); BAX(35); BCR(35); BDX(35); BMS(35); BSX(35);
CAH(35); CFN(35); CI(35); CVH(35); DGX(35); DVA(35); FRX(35); HSP(35);
HUM(35); JNJ(35); LH(35); LLY(35); MCK(35); MDT(35); MKC(35); MRK(35);
PFE(35); PKI(35); STJ(35); SYK(35); THC(35); TMO(35); UNH(35); VAR(35);
WAT(35); WLP(35); AFL(40); AIG(40); AIZ(40); ALL(40); AXP(40); BAC(40);
BBT(40); BEN(40); BK(40); BTU(40); C(40); CB(40); CBG(40); CMA(40); COF(40);
FHN(40); GNW(40); GS(40); HIG(40); JPM(40); KEY(40); L(40); LM(40); LNC(40);
LUK(40); MET(40); MMC(40); MTB(40); NBR(40); NYX(40); PGR(40); PNC(40);
RF(40); SCHW(40); STI(40); STT(40); TMK(40); TRV(40); TSN(40); UNM(40);
USB(40); WFC(40); WM(40); XL(40); A(45); AMD(45); CSC(45); EMC(45); FCX(45);
FIS(45); GLW(45); GME(45); HPQ(45); HRS(45); IBM(45); JBL(45); JNPR(45);
MA(45); MWV(45); TER(45); TSS(45); XRX(45); PCS(50); S(50); T(50); VZ(50);
AEE(55); AEP(55); AES(55); CMS(55); CNP(55); D(55); DTE(55); DUK(55);
ED(55); EIX(55); EQT(55); ETR(55); EXC(55); FE(55); GAS(55); NEE(55);
NI(55); NU(55); PCG(55); PEG(55); PNW(55); POM(55); PPL(55); SCG(55);
SRE(55); TE(55); TEG(55); TXT(55); WEC(55); XEL(55).

\end{document}